\documentclass[prb,amsfonts,amssymb,floats,groupedaddress,superscriptaddress,twocolumn,aps,floatfix,longbibliography,10pt]{revtex4-2}

\usepackage{graphicx}
\usepackage{graphics}
\usepackage{amsmath}
\usepackage{amssymb}
\usepackage{amsfonts}
\usepackage{dsfont}
\usepackage{color}
\usepackage[mathscr]{euscript}
\definecolor{darkblue}{rgb}{0, 0, 0.8}
\usepackage[colorlinks=true, breaklinks=true, linkcolor=red, citecolor=blue, urlcolor=blue]{hyperref} 
\usepackage{hyperref}
\usepackage{subfigure}
\usepackage{xfrac}
\usepackage{bm}
\usepackage{kantlipsum}
\usepackage{enumitem}
\usepackage{tikz}
\usepackage{framed}
\usepackage{graphicx}
\usepackage{subfigure}
\usepackage{cleveref}
\usepackage{array}
\usepackage{nicefrac}
\usepackage{hhline}

\usepackage{orcidlink}

\newcommand{\parL}[1]{\noindent\textbf{\textit{#1}}---}

\newcommand{\ket}[1]{\ensuremath{|#1\rangle}}

\newcommand{\bralket}[2]{\ensuremath{\langle#1|#2\rangle}}

\allowdisplaybreaks[1]

\newcommand{\diagram}[1]{\vcenter{\hbox{\includegraphics[scale=0.75]{tns/#1.pdf}}}}

\newcommand{\code}[1]{\texttt{#1}}

\DeclareMathOperator{\tr}{tr}

\newcommand{\e}{\ensuremath{\mathrm{e}}}

\newcommand{\ham}{\ensuremath{\hat{H}}}

\newcommand{\ulbaddress}{Center for Nonlinear Phenomena and Complex Systems, Universit\'e Libre de Bruxelles, CP 231, Campus Plaine, 1050 Brussels, Belgium}
\newcommand{\solvayaddress}{International Solvay Institutes, 1050 Brussels, Belgium}

\usepackage{ulem}

\begin{document}

\title{Corner entanglement scaling with projected entangled pair states}

\author{Chloé Van Bastelaere\orcidlink{0009-0002-6164-6481}}
\email{chloe.van.bastelaere@ulb.be}
\affiliation{\ulbaddress}
\affiliation{\solvayaddress}

\author{Rui-Zhen Huang}
\email{huangrzh@icloud.com}
\affiliation{Graduate School of China Academy of Engineering Physics, Beijing 100193, China}
 
\author{Laurens Vanderstraeten\orcidlink{0000-0002-3227-9822}}
\email{laurens.vanderstraeten@ulb.be}
\affiliation{\ulbaddress}

\date{\today}

\begin{abstract}
Entanglement scaling provides a powerful probe of universal properties at quantum critical points. In two dimensions, contributions originating from a corner-shaped bipartition exhibit a universal scaling, which is determined by the underlying conformal field theory. We develop a method to extract this corner entanglement entropy from projected entangled pair states directly in the thermodynamic limit. When applied to models at a quantum critical point, we show that the corner contribution exhibits scaling with the effective correlation length, in agreement with the hypothesis of finite-entanglement scaling. Our results for the corner coefficients are consistent with other methods, demonstrating the efficiency of our method for diagnosing strongly-correlated quantum critical points in two dimensions.
\end{abstract}

\maketitle

\parL{Introduction}
%
Quantum entanglement is a central concept in modern condensed matter physics and provides a powerful framework to understand strongly correlated many-body systems.
Over the years, it has become an indispensable tool for detecting and characterizing quantum phases of matter and quantum phase transition.
One particular interest is the access to universal properties of critical systems. Indeed, insights on the underlying conformal field theory (CFT) can be extracted through various entanglement measures.

For one-dimensional gapless systems, the entanglement entropy scales as the logarithm of the length $L$ of the bipartition with a prefactor proportional to the central charge of the underlying CFT~\cite{calabrese:04}.
For higher dimensional systems, such universal properties are also encoded in the entanglement entropy but rather appear in the subleading contributions~\cite{casini:07,bueno:15-2}. 
For instance, for two-dimensional systems at a quantum critical point, the scaling of the entanglement entropy is expected to take the form
\begin{equation}\label{eq:scalingentropy}
 	S = a L + c \log L +\dots\;.
\end{equation}
The leading term is the area law~\cite{eisert:10,wolf:08}, which does not a priori contain any universal information.
In contrast, the subleading logarithmic contribution encodes information about the underlying critical theory~\cite{bueno:15,bueno:15-2}. This correction appears when the boundary of the bipartition contains a corner of opening angle $\theta$. Its coefficient is universal and depends on the opening angle and the universality class of the critical theory.
The corner contribution therefore offers direct access to the universal properties of critical systems.

Naturally, several numerical approaches have been developed to compute the corner contribution in microscopic lattice systems, including quantum Monte Carlo (QMC) methods~\cite{humeniuk:12,helmes:14,helmes:15,ngai:26,song:25,daliao:25,demidio:24,torlai:24}, series expansion~\cite{singh:12,trithep:14}, or numerical linked cluster expansions (NCLE)~\cite{kallin:13,kallin:14,stoudenmire:14,sahoo:16}. Interestingly, for lattice models with $O(N)$ phase transitions, it has been observed that the corner coefficient approximately scales with $N$~\cite{kallin:13,stoudenmire:14,kallin:14}. The corner contribution has also been exploited to diagnose deconfined quantum critical points~\cite{song:25,daliao:23,demidio:24}. Nonetheless, the reliable extraction of the subleading scaling of the corner contribution with system size is very challenging in practice.

Tensor network methods provide a complementary approach for this problem, as they offer an efficient representation of the entanglement structure of many-body quantum states~\cite{cirac:21}.
Indeed, they naturally fulfill the area law, making them a natural framework for modeling ground states of strongly correlated many-body states. As such, tensor networks have become the state-of-the-art numerical approach for simulating one-dimensional systems and are progressively becoming one of the standard tools for simulating ground state properties of two-dimensional systems.

To describe scaling behavior near critical points with tensor networks, the theory of finite-entanglement scaling~\cite{Nishino1996, tagliacozzo:08, pollmann:09, Pirvu2012} has been developed. Although tensor networks are formulated directly in the thermodynamic limit and therefore do not exhibit finite-size effects, truncating the entanglement always introduces a relevant perturbation in the system. This gives rise to an effective correlation length $\xi$, which appears in the scaling laws that contain universal information. For 1$d$ systems, finite-entanglement scaling with matrix product states (MPS) provides an efficient formalism for determining universal signatures and establishing connections with effective field theory descriptions~\cite{tagliacozzo:08,pollmann:09,vanhecke:19, huang:24,Schneider2025}.
A similar program is under development in two dimensions with projected entangled-pair states (PEPS)~\cite{corboz:18,rader:18,vanhecke:22}, allowing for the simulation of quantum critical points of (2+1)-$d$ theories with great accuracy. Nonetheless, the scaling of entanglement entropies in PEPS has not been studied (however, see Ref.~\onlinecite{tagliacozzo:09} for an early work using tree tensor networks).

In this work, we develop a method to extract the universal scaling of the corner contribution to the Rényi-$n$ entropy for ground states of interacting models in 2$d$ using PEPS. In particular, we consider the second Rényi entropy, $n=2$, and focus on a corner with an opening angle of $\theta=\pi/2$. Given a PEPS wavefunction, our method allows for the extraction of the corner entanglement entropy independently from the area law and directly in the thermodynamic limit. We investigate the scaling of the corner entropy with the correlation length for three different cases: the critical Ising model, the critical bilayer XY model, and a gapped chiral spin liquid.

\parL{Methods}\label{sec:peps}
%
%
We consider an infinite PEPS wavefunction parametrized by a single tensor $a$, 
\begin{equation}\label{eq:bipartiton}
    \ket{\Psi(a)} = \diagram{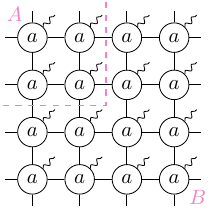} \;.
\end{equation}
The four virtual legs of the tensor have a bond dimension of $D$, controlling the amount of entanglement in the PEPS wavefunction.
For a given model Hamiltonian, we variationally optimize the tensor $a$ using a gradient-based energy minimization~\cite{vanderstraeten:16,corboz:16} to find an optimal approximation for the ground state. During the optimization, the energy is evaluated using approximate contraction techniques with an environment bond dimension $\chi_E$.

Given the PEPS wavefunction, we consider a bipartition of the state with subregions $A$ and $B$ and focus on the Rényi-$n$ entropy, $n>1$, defined as
\begin{equation}
	S_n = \frac{1}{1-n} \log \tr \rho_A^n,
\end{equation}
where $\rho_A$ is the reduced density matrix of the subregion $A$ of the system.
The entanglement spectrum from a PEPS can be accessed using the bulk-boundary correspondence~\cite{cirac:11}. This method allows us to write an operator $\tilde{\rho}$ that is isospectral to $\rho_A$, as 
\begin{equation}\label{eq:bulkboundary}
	\tilde{\rho} = M^AM^B,
\end{equation} 
where $M^{A,B}$ are the effective environments obtained from contracting the subregions $A$ and $B$. These environments can typically be approximated efficiently by tensor-network contraction techniques. This gives us then an efficient method for computing entanglement spectra and entropies.

We first consider a linear bipartition. Using the VUMPS algorithm~\cite{zaunerstauber:18, fishman:18, vanderstraeten:19}, we approximate the linear environment as a boundary MPS, which is parametrized by the tensor $M$ with bond dimension $\chi$.
From the bulk-boundary correspondence, we find that the linear reduced density matrix is given by
\begin{equation}\label{eq:rdm_linear}
    \Tilde{\rho}_{\ell} = \cdots \diagram{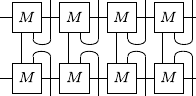} \cdots \;, 
\end{equation}
which we normalize as $\tr\Tilde{\rho}_{\ell}=1$ by rescaling the tensor $M$. For simplicity, we assume that the environments of both subregions can be approximated by the same boundary MPS.
Similarly, we find the reduced density matrix squared, which allows us to compute the Rényi-2 entropy as the infinite power of a four-layer channel operator:
\begin{equation}\label{eq:area}\begin{aligned}
    S_{\ell, 2} = -\log \left(\cdots\diagram{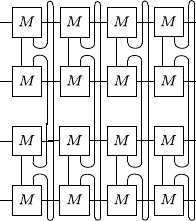}\cdots\right) = -\log(\lambda) \; L,
\end{aligned}\end{equation}
where $\lambda$ is the leading eigenvalue of this channel operator and $L\to\infty$ is the linear size of the bipartition. We recover exactly the expression of the area law~\eqref{eq:scalingentropy} with coefficient $a_2=-\log\lambda$ directly in the thermodynamic limit.

Let us now consider a corner-shaped bipartition of length $L=L_x+L_y$, as represented in~\eqref{eq:bipartiton}.
We construct the boundary MPS for such a geometry following the approach developed for channel environments methods~\cite{vanderstraeten:15,vanderstraeten:16}.
Away from the corners, the boundary MPSs are well approximated by the ones of the linear problem $M$.
The boundary MPS of the corner-shaped region $A$ is then obtained by inserting the corner tensor $V$, whereas the one for the region $B$ can be captured by a tensor $V'$,
\begin{equation}
    M^A = \diagram{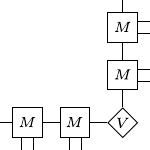}\;, \; M^B=\diagram{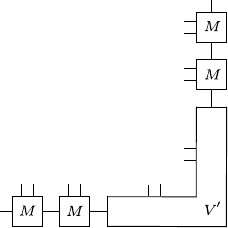}\;.
\end{equation}
These corner tensors can be computed by solving fixed-point equations~\cite{vanderstraeten:15,vanderstraeten:16}.
For clarity, we will flatten the diagram in the following such that the tensors on the left (resp. right) of a corner tensor correspond to the horizontal (resp. vertical) direction.
With these expressions for the boundary MPSs, we use the bulk-boundary correspondence~\eqref{eq:bulkboundary} to compute Rényi-2 entanglement entropy.
The reduced density matrix is
\begin{equation} \label{eq:rho_c}
    \Tilde{\rho}_{c} = \cdots \diagram{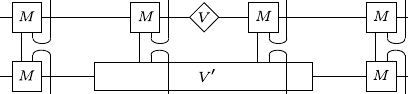}\cdots,
\end{equation}
where the left (resp. right) semi-infinite channels are repeated $L_x-1$ (resp. $L_y-1$) times. These channels are exactly the ones found in the case of a linear bipartition~\eqref{eq:rdm_linear}. We normalize the corner tensors such that $\tr\Tilde{\rho}_{c}=1$. 
From this expression of $\rho_{c,A}$, we compute the reduced density matrix squared and find
\begin{equation}
    \tr \Tilde{\rho}_{c}^2 = \lambda^{L_x+L_y-2} \Lambda,
\end{equation}
with 
\begin{equation}\label{eq:rdm2_corner}
    \Lambda = \diagram{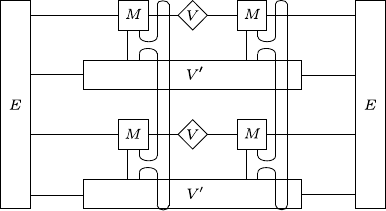} \;.
\end{equation}
where $E$ is the fixed point of the channel operator that appeared in the expression for the linear entropy~\eqref{eq:area}.
This allows us to write the Rényi-2 entropy of a corner-shaped region as
\begin{equation}
    S_{2} = - (L_x+L_y)\log \lambda -  \log \left(\frac{\Lambda}{\lambda^2}\right) = S_ {\ell,2} + S_{c,2}.    
\end{equation}
We observe that the expression is composed of the area law $S_{\ell,2}$~\eqref{eq:area} and an additional term $S_{c,2}$ which we identify to be the corner contribution. This contribution can be computed independently from the area law by contracting the tensor~\eqref{eq:rdm2_corner}.

By comparison with the finite-size scaling of the corner contribution in~\eqref{eq:scalingentropy}, $S_{c,2}$ should scale logarithmically with the effective length scale of the system. According to the hypothesis of finite-entanglement scaling for PEPS simulations~\cite{rader:18, corboz:18, vanhecke:22}, the relevant length scale in a PEPS wavefunction is the correlation length $\xi$, which can be extracted from the leading eigenvalue of the transfer matrix of the linear boundary MPS. We therefore expect that $S_{c,2}$ scales as $c_2 \log(\xi)$.
To find this behavior and extract $c_2$, we proceed as follows.
For a given Hamiltonian, we optimize a PEPS tensor with bond dimension $D$ and (relatively high) $\chi_E$. From this state, we compute the boundary MPS $M$ for a large range of $\chi$.
For each $\chi$, we calculate the area law prefactor $a_2$, the corner contribution $S_{c,2}$ and the correlation length $\xi$. We then obtain the corner coefficient $c_2$ as the slope of $S_{c,2}$ as a function of $\log(\xi)$.
In the Appendix~\ref{app:entropy}, we provide a detailed review of the bulk-boundary correspondence and a detailed derivation of the corner-shaped entanglement entropy with PEPS.
%
%
%

\parL{Results}\label{sec:benchmark}
%
As a first application, we use our method to extract the corner entropy contribution for the critical 2$d$ transverse field Ising model (TFIM), which is described by the Hamiltonian 
\begin{equation}\label{eq:ising hamiltonian}
    \ham = -\sum_{\langle ij\rangle} \sigma^z_i \sigma^z_j - g \sum_{i} \sigma^x_i.
\end{equation}
This model exhibits a phase transition at $g=g_c \approx 3.04438$~\cite{blote:02}, going from a symmetry-broken phase to a polarized phase, belonging to the 3$d$ Ising universality class.
For this model, the corner contribution of the Rényi-2 entropy has been computed with various methods such as QMC~\cite{humeniuk:12, inglis:13, ngai:26}, NLCE~\cite{kallin:13,sahoo:16} or series expansions~\cite{singh:12}. The different results from the existing literature are gathered in Table~\ref{tab:valuesising}.
\begin{table}
    \begin{tabular}{|l||l|l|}
        \hline
        Method & TFIM & Bilayer XY \\
        \hline
        QMC  & $\;-0.0075(25)$~\cite{humeniuk:12} & $\;-0.010(2)$~\cite{helmes:15} \\
          & $\;-0.006(2)$~\cite{inglis:13} &\\
          & $\;-0.0050(5)$~\cite{ngai:26} &\\
        NLCE  & $\;-0.0053$~\cite{kallin:13} & $\;-0.0111(1)$~\cite{stoudenmire:14}\\
          & $\;-0.0059$~\cite{sahoo:16} &\\
        Series expansion & $\;-0.0055(5)$~\cite{singh:12} & $\;-0.0125(6)$~\cite{trithep:14}\\
        \hline
        \textbf{PEPS (our work)} & $\;-0.0050(1)$ & $\;-0.012(2)$\\
        \hline
    \end{tabular}
    \caption{Comparison of our PEPS results and the literature for the corner coefficient $c_2$ of the transverse field Ising model and bilayer XY model.}
    \label{tab:valuesising}
\end{table}

We optimize the ground state of the model at bond dimension $D=3$ and $\chi_E = 200$ at the critical point $g_c$. The TFIM has already been studied extensively with PEPS, and the phase transition is already well captured for this value of the bond dimension $D$~\cite{orus:09,vanderstraeten:16,rader:18,corboz:18}.
From this optimized tensor, we calculate the boundary MPS with bond dimension $\chi=5,6,\dots,80$. For each $\chi$, we compute the area law and the corner coefficient following the scheme presented above.
\begin{figure}
    \centering
    \includegraphics[width=\columnwidth]{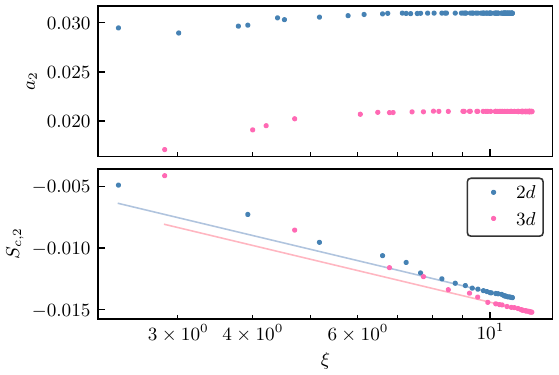}
    \caption{Scaling of (top) the area law prefactor and (bottom) the corner contribution as a function of the correlation length for the 2$d$ TFIM and the 3$d$ classical Ising model at criticality. The dots are the data points, and the continuous lines are the linear fit. For the clarity of the bottom plot, we only show one in three data points.}
    \label{fig:TFIM corner scaling}
\end{figure}

First, we look at the area law contribution. The top panel of Fig.~\ref{fig:TFIM corner scaling} shows the area law prefactor as a function of $\xi$. We observe that $a_2$ rapidly converges to a constant value, and does not scale with the correlation length, as we expect for the area law.

We now look at the scaling of the corner contribution at the critical point. In the bottom panel of Fig.~\ref{fig:TFIM corner scaling}, we observe the predicted $\log(\xi)$ scaling of the corner contribution, demonstrating that our method captures the expected universal behavior. As this optimized PEPS at finite bond dimension $D$ is known to exhibit a finite correlation length~\cite{rader:18,corboz:18} in the limit $\chi\to\infty$, the scaling law necessarily halts at a certain value $\xi_\infty(D)$.
Using a linear fit, we find $c_2=-0.0050(1)$. This value is close to the range of values already known in the literature and is particularly close to the most recent QMC result~\cite{ngai:26}. However, we note that this value is sensitive to the fitting method used, as discussed in the Appendix~\ref{app:ising}.

To further confirm the universality of this result, we consider the classical 3$d$ Ising model, for which the critical point falls in the same universality class as the TFIM. We find a PEPS wavefunction that approximates the fixed point of the plane-to-plane transfer matrix of the 3$d$ partition function~\cite{vanderstraeten:18}, for which we then extract the corner contribution with the methods introduced above.
We find that the corner contribution again follows a logarithmic law, and we extract the value $c_2=-0.00503(2)$.
The correspondence of the classical and quantum results showcases the universal property of the corner coefficient.

Next, we investigate the bilayer XY model, which is described by the Hamiltonian
\begin{multline}
    \ham_{XY} =  \sum_{\langle i,j\rangle} S^x_{1,i}S^x_{1,j} + S^y_{1,i}S^y_{1,j} +S^x_{2,i}S^x_{2,j} + S^y_{2,i}S^y_{2,j} \\ + J_{\perp} \sum_i S^x_{1,i}S^x_{2,i} + S^y_{1,i}S^y_{2,i}. 
\end{multline}
This model undergoes a phase transition that belongs to the $O(2)$ universality class at $J_{\perp}=5.460(1)$~\cite{helmes:15}, going from an anti-ferromagnetically ordered phase at low $J_{\perp}$ to a dimer-ordered phase at high $J_{\perp}$.
To represent the bilayer structure, we consider a PEPS tensor with a local physical dimension of four, variationally optimize the PEPS for the model at criticality with $D=5$ and $\chi_E=200$ and follow the same procedure as before.

\begin{figure}
    \centering
    \includegraphics[width=\columnwidth]{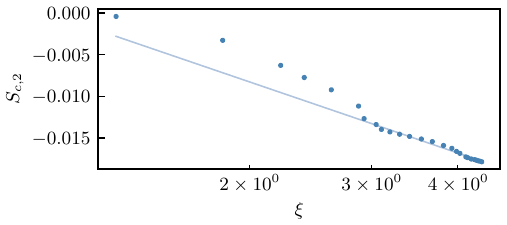}
    \caption{Scaling of the corner contribution as a function of the correlation length for the bilayer XY model at criticality. The dots are the data points, and the continuous line is the linear fit. For the clarity of the plot, we only show one in three data points.}
    \label{fig:Bilayer corner scaling}
\end{figure}

Our results are shown in Fig.~\ref{fig:Bilayer corner scaling}. We find that the corner contribution again scales with $\log(\xi)$, with a fitted prefactor of $c_2=-0.012(2)$. This result is close to two times the value obtained for Ising, which is what we expect for a phase transition in the $O(2)$ universality class. Within error bars, it also matches the results found through various other methods (see Table~\ref{tab:valuesising}).

\begin{figure}[t!]
    \centering
    \includegraphics[width=\columnwidth]{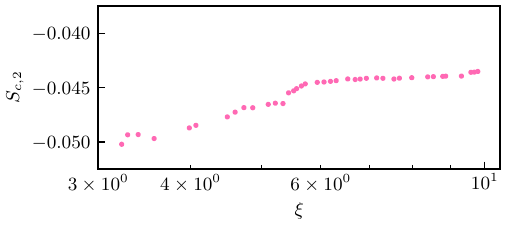}
    \caption{Scaling of the corner contribution $S_{c,2}$ as a function of the correlation length for a gapped chiral spin liquid for $D=5$.}
    \label{fig:CSL}
\end{figure}

As a final application, we study a chiral spin liquid (CSL) state that was obtained variationally in Ref.~\onlinecite{hasik:22}. This is a gapped state, and the corner contribution is not expected to scale logarithmically~\cite{estienne:22}. However, a PEPS approximation is believed to always host critical correlations~\cite{Wahl2013, Dubail2015}, although the magnitude of these correlations vanishes upon increasing the bond dimension~\cite{hasik:22}. In Fig.~\ref{fig:CSL} we observe that the corner contribution does not scale with $\log(\xi)$ as in our previous examples, but rather converges to a finite value. This confirms that PEPS describe the gapped nature of chiral topological states correctly.

%

\parL{Outlook}
%
In this work, we have developed a method to extract the subleading correction of the entanglement entropy originating from a corner-shaped bipartition in a PEPS wavefunction. 
We have applied this method to variational PEPS approximations for models at a quantum critical point, showing that the corner contribution scales logarithmically with the effective correlation length that is induced by the finite environment bond dimension, in analogy to the known finite-size scaling of Eq.~\ref{eq:scalingentropy}. We believe these results show that finite entanglement scaling in PEPS can be used effectively for extracting universal properties of (2+1)-$d$ CFT at quantum critical points, particularly in the cases where QMC methods suffer from sign problems.

Looking forward, a natural extension is to investigate higher Rényi-$n$ entropies, which provide additional information on the entanglement spectrum. Our method can easily be generalized for $n>2$, but since the size of the tensors $E$ in~\eqref{eq:rdm2_corner} grows exponentially in $n$, further approximations would be needed to evaluate the corner contribution.
We have here only focused on the case of an entanglement region with a single corner in an infinite square lattice. It would also be interesting to adapt our method to different lattice geometries and broader ranges of opening angles~\cite{bueno:15-2, estienne:22}, as well as entanglement regions with physical edges and corners~\cite{estienne:20, rozon:20}. Another promising prospect is to investigate the influence of a corner-shaped bipartition on the entanglement spectrum and the associated states~\cite{sirois:21}. Indeed, interpreting the expression in~\eqref{eq:rho_c} as a representation of the corner entanglement Hamiltonian $H_c$ as $\tilde{\rho}_{c}\propto\e^{-H_c}$, the presence of a corner is modeled as an impurity, for which we can adapt existing methods for extracting entanglement spectra from PEPS~\cite{haegeman:17, hasik:22}.

\newpage
\parL{Acknowledgments}%
%
L.V. would like to thank Andreas Läuchli for inspiring discussions. R.Z.H would like to thank Hui-Ke Jin for helpful discussions. C.V. is financially supported by the FRIA Grant FC 63985. R.Z.H is supported by the National Natural Science Foundation of China (Grant No. 12504183). Computational resources have been provided by the Consortium des Equipements de Calcul Intensif (CECI), funded by the F.R.S.-FNRS under Grant No. 2.5020.11 and by the Walloon Region. Tensor network simulations are implemented using TensorKit~\cite{TensorKit} and PEPSKit~\cite{PEPSKit}.

\bibliography{bibliography.bib}
\newpage
\onecolumngrid
\appendix
%
\section{Details on the entanglement entropy calculation}\label{app:entropy}
%
In this appendix, we give technical details on the derivation of the corner-shaped entanglement entropy for a PEPS wavefunction. We first review the PEPS bulk-boundary correspondence~\cite{cirac:11}, then the case of a linear bipartition, and finally we derive the expression for the corner contribution.
\subsection{Bulk-boundary correspondence}\label{app:bulk}
We consider an infinite PEPS state defined by the tensor $a$,
\begin{equation}
    |\Psi\rangle = \diagram{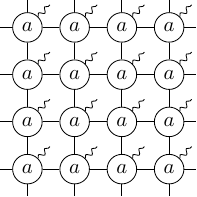}\;.
\end{equation}
We consider a bipartition with a general geometry of this state with the subregions $A$ and $B$.
The state can be written as
\begin{equation}
    |\Psi\rangle = \sum_{\{\gamma_i\}} |\psi^A_{\{\gamma_i\}}\rangle \otimes |\psi^B_{\{\gamma_i\}}\rangle,
\end{equation}
where ${\{\alpha_i\}}$ denote the set of virtual bonds on the boundary of the bipartition and $|\psi^{A,B}\rangle$ are the semi-infinite states representing each subregion.
The only open indices are then the physical and the boundary ones.
We make a decomposition of these states as an isometry $U_{A,B}$ between the physical degrees of freedom and the virtual ones, and an operator $V_{A,B}$ acting on the virtual boundary ones only. For the subregion $A$, the state is then written is
\begin{equation}
    \ket{\psi^A_{\{\gamma_i\}}} = |U_A^{\{\alpha_i\}}\rangle V^{\{\alpha_i\},\{\gamma_i\}}_A,
\end{equation}
where we have omitted the explicit sums over repeated $\alpha$ indices. The isometrical property of the states $\ket{U_A}$ is
\begin{equation}
    \bralket{U_A^{\{\alpha'_i\}}}{U_A^{\{\alpha_i\}}} = \delta_{\{\alpha'_i\},\{\alpha_i\}}.
\end{equation}
Similarly, for the subregion $B$, we decompose the state as
\begin{equation}
    \ket{\psi^B_{\{\gamma_i\}}} = V_B^{\{\gamma_i\},\{\beta_i\}}\ket{U_B^{\{\beta_i\}}}. 
\end{equation}
With these decompositions, the PEPS wavefunction is now given by
\begin{equation}
    |\Psi\rangle =  V_A^{\{\alpha_i\},\{\gamma_i\}} V_B^{\{\gamma_i\},\{\beta_i\}} 
    \ket{U_A^{\{\alpha_i\}}} \otimes  \ket{U_B^{\{\beta_i\}}}.
\end{equation}
The reduced density matrix of the subregion $A$, $\rho_A= \tr_B|\Psi\rangle\langle\Psi|$, is then expressed as
\begin{equation}
\begin{aligned}
    \rho_A &= V_A^{\{\alpha_i\},\{\gamma_i\}} V_B^{\{\gamma_i\},\{\beta_i\}} \overline{V}_B^{\{\beta'_i\},\{\gamma'_i\}}\overline{V}_A^{\{\gamma'_i\},\{\alpha'_i\}}|U^{\{\alpha_i\}}_A\rangle \langle U^{\{\alpha'_i\}}_A|\delta_{\{\beta'_i\},\{\beta_i\}}.
\end{aligned}
\end{equation}
Moreover, since $U_{A}$ is an isometry, we define
\begin{equation}
\begin{aligned}
    \Tilde{\rho}_A^{\{\alpha'_i\},\{\alpha_i\}} &= \langle U_A^{\{\alpha'_i\}}|\rho_A|U_A^{\{\alpha_i\}}\rangle \\
    &= V_A^{\{\alpha_i\},\{\gamma_i\}} V_B^{\{\gamma_i\},\{\beta_i\}} \overline{V}_B^{\{\beta'_i\},\{\gamma'_i\}}\overline{V}_A^{\{\gamma'_i\},\{\alpha'_i\}}\delta_{\{\beta'_i,\beta_i\}},
\end{aligned}
\end{equation}
which has the same spectrum as $\rho_A$ and only acts on the virtual boundary degrees of freedom. We then have that
\begin{equation}
    \mathrm{spec}\left(\rho_A\right)=\mathrm{spec}\left(\Tilde{\rho}_A\right) = \mathrm{spec}\left(\overline{V}_AV_AV_B\overline{V}_B\right).
\end{equation}
The operators $V_{A,B}$ can be found as the square root of the half-infinite contraction which is well approximated by the boundary MPS $M^{A,B}$. Indeed, we compute
\begin{equation}
    \bralket{\psi^A_{\{\gamma_i'\}}}{\psi^A_{\{\gamma_i\}}} = V_A^{\{\alpha_i\},\{\gamma_i\}} \overline{V}_A^{\{\gamma_i'\},\{\alpha_i\}} = M^{A^{\{\gamma_i'\},\{\gamma_i\}}},
\end{equation}
and similarly
\begin{equation}
\begin{aligned}
   M^{B^{\{\gamma_i'\},\{\gamma_i\}}} = \bralket{\psi^B_{\{\gamma_i'\}}}{\psi^B_{\{\gamma_i\}}} = V_B^{\{\gamma_i\},\{\beta_i\}} \overline{V}_B^{,\{\beta_i'\},\{\gamma_i'\}} \delta_{\{\beta_i'\},\{\beta_i\}}.
\end{aligned}
\end{equation}
The spectrum of the reduced density matrix is then given by
\begin{equation}
    \mathrm{spec}\left(\rho_A\right) = \mathrm{spec}\left(M^AM^B\right).
\end{equation}
This expression can be generalized as
\begin{equation}
    \mathrm{spec}\left(\rho_A^n\right) = \mathrm{spec}\left(\left(M^AM^B\right)^n\right).
\end{equation}
%
%
\subsection{Linear bipartition}\label{app:linear}
%
%
Let us consider first a linear bipartition of our system, i.e. a line of length $L$. We consider a PEPS with rotation and reflection symmetries imposed, 
\begin{equation}
    \diagram{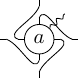} \;= \;\diagram{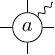}\;, \quad \diagram{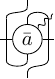} \;= \;\diagram{PEPS}\;.
\end{equation}
Following the bulk-boundary correspondence, the entanglement spectrum of the PEPS wavefunction is given by the leading eigenvector or fixed point of the PEPS transfer matrix $T$,
\begin{equation}
    \diagram{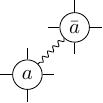}= \;\diagram{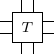}\;.
\end{equation}
This fixed-point is well approximated by a boundary MPS $M$ which is defined through the equation
\begin{equation}\label{eq:linearcontraction}
    \diagram{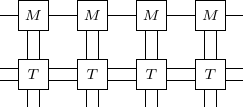}\; \propto \;\diagram{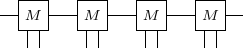}\;.
\end{equation}
We find the leading eigenvector of $T$ as a MPS with bond dimension $\chi$ using the VUMPS algorithm~\cite{zaunerstauber:18,fishman:18,vanderstraeten:19}. With the symmetries considered here, the tensor $M$ defines the boundary MPS of both $A$ and $B$.
The trace of the reduced density matrix is then given by
\begin{equation}
    \begin{aligned}
        \tr\rho_{\ell,A} = \tr\left(\;\diagram{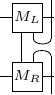}\;\right)^L = \omega^L,
    \end{aligned}
\end{equation}
where $M_{L,R}$ denote the left and right gauge of $M$ and $\omega$ is the leading eigenvalue of the fixed-point equation
 \begin{equation}\label{eq:trace_rdm}
 	\diagram{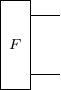}\; \propto \diagram{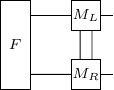}\;, \qquad \diagram{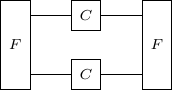} = 1,
\end{equation}
and with $C$ the center gauge tensor of $M$.
Furthermore, the trace of the reduced density matrix squared is given by
\begin{equation}
\begin{aligned}
    \tr\rho_{\ell,A}^2 = \tr\left(\;\diagram{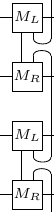}\;\right)^L=\lambda^L,
\end{aligned}
\end{equation}
where $\lambda$ is the leading eigenvalue of the fixed-point equation
\begin{equation}
	\diagram{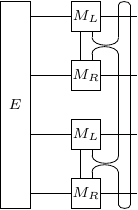}\; \propto \diagram{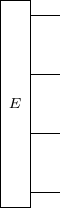}\;, \qquad \diagram{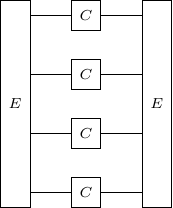} = 1.
\end{equation}
We then find that the Rényi-2 entropy is given by
\begin{equation}
    S_{\ell, 2} = -\log\left(\frac{\tr\rho^2_{\ell,A}}{\left(\tr\rho_{\ell,A}\right)^2}\right) = - \log\left(\frac{\lambda}{\omega^2}\right) L.
\end{equation}
Here, the denominators are used as a normalization condition. We recognize exactly the expression of the area law with $a_2= - \log\left(\frac{\lambda}{\omega^2}\right)$.
\subsection{Corner-shaped bipartition}\label{app:corner}
We now consider a corner-shaped bipartition,
\begin{equation}\label{eq:appbipartition}
    |\Psi\rangle = \diagram{Bipartition}\;.
\end{equation}
We consider a bipartition of length $L=L_x+L_y$. We again follow the bulk-boundary correspondence and compute the entanglement spectrum from fixed-point of the PEPS transfer matrix that now has a specific geometry.
Let us first review a previous corner-shaped tensor network algorithm~\cite{vanderstraeten:15,vanderstraeten:16} on which our method is based on. 
The central idea is to, instead of considering the linear transfer matrix as in~\eqref{eq:linearcontraction}, introduce a corner-shaped transfer matrix. We again approximate the fixed point of this transfer matrix by a boundary MPS. Far from the corner, the boundary MPS is well approximated by the one of the linear problem $M$. To take into account the corner-shaped geometry of the transfer matrix, we introduce a corner tensor $V$ which is computed through a fixed-point equation,
\begin{equation}\label{eq:appcorner1}
	\diagram{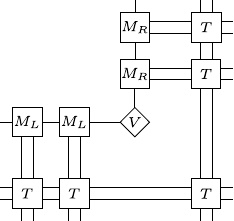}\; \propto \; \diagram{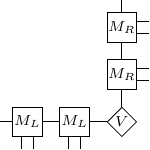}\;,
\end{equation}
Applying this method on each direction of a 2$d$ lattice, we can calculate the norm of the PEPS wavefunction as
\begin{equation}\label{eq:normchannel}
    \diagram{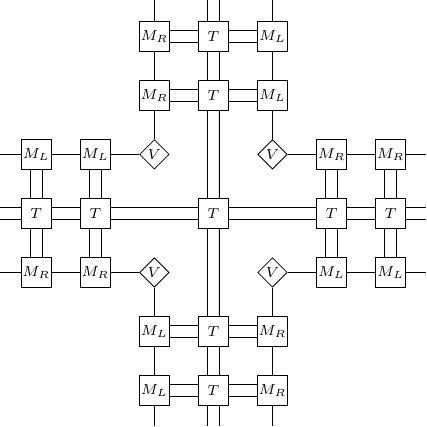}\; \propto \; \diagram{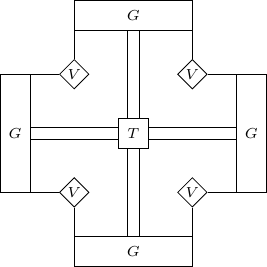}\;,
\end{equation}
where the tensors $G$ are the fixed point of the channel operators
\begin{equation}
    \diagram{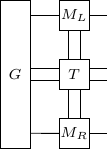}\; \propto \;\diagram{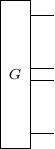}\;.
\end{equation}
With this method in mind, let us now focus on the entanglement spectrum. As shown in~\eqref{eq:appbipartition}, the subregion $A$ has exactly the shape of the environment used in the contraction method~\eqref{eq:appcorner1}. Its boundary MPS is then given by
\begin{equation}
    M^A = \diagram{CornerTensor_1}\;.
\end{equation}
To construct the boundary MPS of $B$, we combine the south and east channels shown in~\eqref{eq:normchannel} to obtain
\begin{equation}
	M^B = \diagram{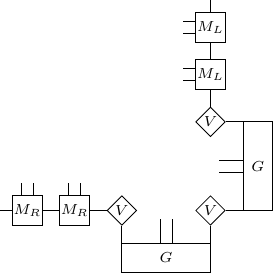}\;.
\end{equation}
With these corner-shaped boundary MPSs, we compute reduced density matrices using the relation~\eqref{eq:bulkboundary}. For the clarity of the diagrams, we will flatten the corner-shaped boundary MPSs in the following so that the tensors on the left (resp. right) of a corner tensor correspond to the horizontal (resp. vertical) direction.

For Rényi entropies, we are interested in traces of reduced density matrices of the corner-shaped subregions. We have that
\begin{equation}
\begin{aligned}
	\tr \rho_{c,A} &= \tr \left( \cdots\; \diagram{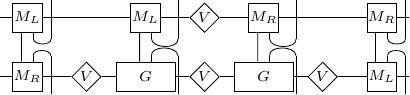}\; \cdots\right)= \omega^{L_x+L_y-2} \;\diagram{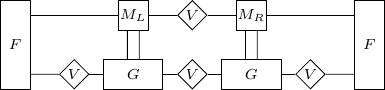}\;.
\end{aligned}
\end{equation}
We see that we recover the same semi-infinite channels as in the linear case repeated $L_x-1$ times on the left and $L_y-1$ times on the right.
Furthermore, the trace of the reduced density matrix squared is
\begin{equation}
\begin{aligned}
	\tr \rho_{c,A}^2 &= \tr \left( \cdots \; \diagram{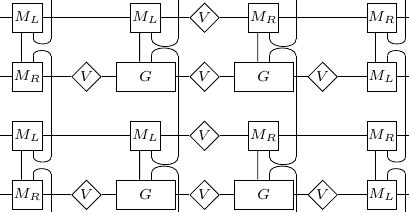}\; \cdots\right) = \lambda^{L_x+L_y-2}\; \diagram{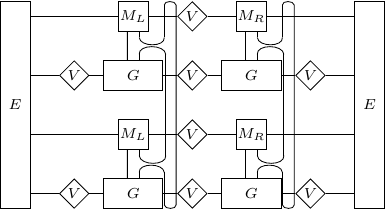}\;.
\end{aligned}
\end{equation}
We find again the same semi-infinite channel as in the linear case and therefore the leading eigenvector $E$.
The expression of the Rényi-2 entropy is then given by, with the proper normalization,
\begin{equation}
\begin{aligned}
	S_2 &= - \log \left(\frac{\tr\rho^2_{c,A}}{\left(\tr\rho_{c,A}\right)^2}\right)\\
	& = -\log\left(\frac{\lambda}{\omega^2}\right) L - \log\left(\frac{\omega^4}{\lambda^2}\right)- \log\frac{\diagram{RDM2_CornerTrace}}{\left(\;\diagram{RDM_CornerTrace}\;\right)^2}\;.
\end{aligned}
\end{equation}
We see that the first term of $S_2$ is exactly the area law found as found for the linear bipartition.
By comparison with the scaling of the entanglement, we identify the rest of the expression as the corner contribution:
\begin{equation}
\begin{aligned}
	S_{c,2} &= - \log\left(\frac{\omega^4}{\lambda^2}\right) - \log\frac{\diagram{RDM2_CornerTrace}}{\left(\;\diagram{RDM_CornerTrace}\;\right)^2}\;.
\end{aligned}
\end{equation}
%
%
\section{Details on the results}
%
%
In this Appendix, we report additional details on the results for the different models used as applications in the main text.
%
%
\subsection{Transverse field Ising model}\label{app:ising}
%
The first model we investigate in the transverse field Ising model, described by the Hamiltonian
\begin{equation}
    \ham = -\sum_{\langle ij\rangle} \sigma^z_i \sigma^z_j - g \sum_{i} \sigma^x_i.
\end{equation}
This model undergoes a phase transition at $g_c=3.04438$. We optimize a PEPS tensor at criticality with $D=3$ and $\chi_E=200$.
%
%
%
\begin{figure}[t!]
    \centering
    \includegraphics[width=0.49\linewidth]{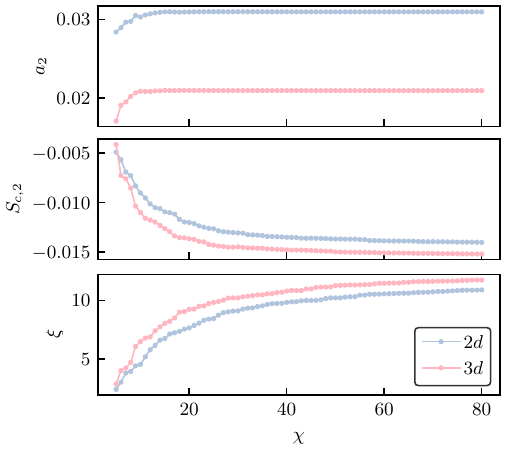}
    \includegraphics[width=0.49\linewidth]{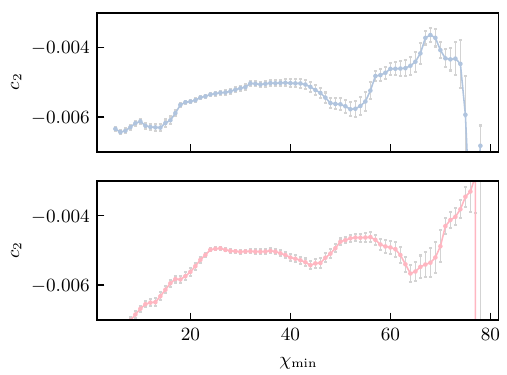}
    \caption{(Left) The area law prefactor $a_2$, the corner contribution $S_{c,2}$, and the correlation length $\xi$ as a function of the bond dimension $\chi$ of the boundary MPS for the $2d$ quantum and the $3d$ classical Ising model at criticality. The continuous lines are a guide for the eye. (Right) Corner coefficient $c_2$ extracted for the 2$d$ TFIM (top) and the 3$d$ classical Ising model (bottom) for different window sizes.}
    \label{fig:isingdata}
\end{figure}
Following the method presented above, we extract the area law coefficient $a_2$, and the corner contribution $c_2$. To this end, we compute the boundary MPS with bond dimension $\chi=5,6,\dots,80$. In Fig.~\ref{fig:isingdata}, we see that the area law coefficient $a_2$ rapidly converges as a function of $\chi$.
We also show the full corner contribution $S_{c,2}$ and the correlation length $\xi$ as a function of $\chi$. To extract the corner coefficient $c_2$, we plot $S_{c,2}$ as a function of $\log\xi$. As we expect from a critical theory, the relation is linear, and $c_2$ corresponds to the slope, see Fig.~\ref{fig:TFIM corner scaling}.
We note, however, that there are some discrepancies between the results obtained from different linear fitting windows, which make the extraction of $c_2$ challenging.
Indeed, the value $c_2$ obtained through a linear fit is sensitive to the choice of the fitting window. Although the values we obtain are systematically close to the results reported by other methods (see Table~\ref{tab:valuesising}), extracting a single value with high precision remains a challenge. Similar difficulties have also been encountered with NCLE and QMC methods.
Here, the difficulties are twofold. On the one hand, the data at small $\chi$ are typically not physically accurate, as commonly observed in tensor network simulations. On the other hand, as the PEPS wavefunction has a finite correlation length, the data points are not equally spaced in $\log\xi$ and the largest-$\chi$ are strongly clustered. As a consequence, small numerical variations in these data points have a significant impact on the fitted slope.

To obtain a robust result, we study the influence of the fitting window on the extracted coefficient. We perform linear fits using the data in the interval $[\chi_{\min},80]$ and vary $\chi_{\min}$. 
As shown in Fig.~\ref{fig:isingdata}, a plateau forms at $c_2\approx -0.005$ for $\chi_{\min}\approx 30,\dots, 44$. This stable regime suggests that the intermediate $\chi_{\min}$ data provide a good estimate of $c_2$. For larger values of $\chi_{\min}$, the extracted coefficient becomes unstable. This instability is explained by the reduced range of $\log\xi$ and the strong clustering of the largest-$\chi$ data points, making the slope sensitive to tiny numerical fluctuations. We consider a data point in this plateau as our result and its associated error bars. We find $c_2=-0.0050(1)$ for $\chi_{\min}=39$. We note that these error bars are associated with the fitting function and not the physics itself.
Moreover, we note that this value matches the most recent QMC method~\cite{ngai:26} with high precision.
Using the same fitting method for the 3$d$ classical model, we obtain that the corner coefficient also forms a plateau at $c_2\approx-0.005$ for intermediate $\chi_{\min}\approx 25,\dots,38$, see Fig.~\ref{fig:isingdata}. We report $c_2=-0.00503(2)$ at $\chi_{\min}=35$ as our result.
%
%
%
%
\subsection{Bilayer XY model}
\begin{figure}[t!]
    \centering
    \includegraphics[width=0.49\linewidth]{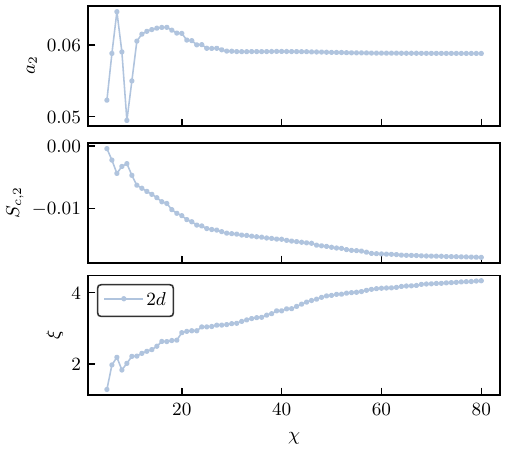}
    \includegraphics[width=0.49\linewidth]{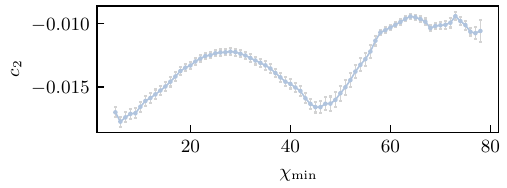}
    \caption{(Left) The area law prefactor $a_2$, the corner contribution $S_{c,2}$, and the correlation length $\xi$ as a function of the bond dimension $\chi$ of the boundary MPS for the bilayer XY model at criticality. The continuous lines are a guide for the eye. (Right) Corner coefficient $c_2$ extracted for the bilayer XY model for different window sizes.}
    \label{fig:bilayerdata}
\end{figure}

%
%
The bilayer XY model is described by the Hamiltonian,
\begin{equation}
    \begin{aligned}
    \ham_{XY} &=  \sum_{\langle i,j\rangle} S^x_{1,i}S^x_{1,j} + S^y_{1,i}S^y_{1,j} +S^x_{2,i}S^x_{2,j} + S^y_{2,i}S^y_{2,j}+ J_{\perp} \sum_i S^x_{1,i}S^x_{2,i} + S^y_{1,i}S^y_{2,i}.
    \end{aligned}
\end{equation}
To be able to represent the ground state of this model on a single-site unit cell, we perform a unitary transform so that the Hamiltonian is now described by
\begin{equation}
    \begin{aligned}
    \ham_{XY} &=  \sum_{\langle i,j\rangle} -S^x_{1,i}S^x_{1,j} + S^y_{1,i}S^y_{1,j} -S^x_{2,i}S^x_{2,j} + S^y_{2,i}S^y_{2,j}+ J_{\perp} \sum_i S^x_{1,i}S^x_{2,i} + S^y_{1,i}S^y_{2,i}.
    \end{aligned}
\end{equation}
We represent the bilayer structure on the level of the PEPS tensor by setting a physical bond dimension of four, which represents two flavors of spins (one for the first layer and the other for the second one).
We optimize a PEPS tensor with $D=5$ and $\chi_E=200$ at criticality $J_{\perp}=5.46$. In the right panels of Fig~\ref{fig:bilayerdata}, we show the area law coefficient $a_2$, the corner contribution $S_{c,2}$, and the correlation length $\xi$ as a function of $\chi$. We see again that the area law coefficient converges rapidly to $a_2=0.0588$. To extract the corner coefficient, we proceed in the same way as for Ising models. In the left panel of Fig~\ref{fig:bilayerdata}, we show $c_2$ obtained for different window sizes $[\chi_{\min},80]$. We observe that the largest plateau forms at $c_2\approx -0.012$ for $\chi_{\min}\approx 21,\dots, 33.$ We consider the value in the middle of this plateau as our result and obtain $c_2=-0.012(2)$ for $\chi_{\min}=26$.

\end{document}